\documentclass[preprint,showpacs,preprintnumbers,amsmath,amssymb,floatfix]{revtex4}

\usepackage{graphicx}
\usepackage{dcolumn}
\usepackage{bm}

\begin{document}


\title{Mean Field Calculation of Thermal Properties
of Simple Nucleon Matter on a Lattice}

\author{T. Abe,${}^1$ R. Seki,${}^{2,3}$ and A. N. Kocharian${}^2$}

\affiliation{ ${}^1$ Department of Physics, Tokyo Institute of
Technology,
Megro, Tokyo 152-8551, Japan\\
${}^2$ Department of Physics and Astronomy, California State
University, Northridge, Northridge, CA 91330, USA\\
${}^3$ W. K. Kellogg Radiation Laboratory, California Institute of
Technology, Pasadena, CA 91125, USA}
\date{\today}

\begin{abstract}
Thermal properties of single species nucleon matter are
investigated assuming a simple form of the nucleon-nucleon
interaction.  The nucleons are placed on a cubic lattice, hopping
from site to site and interacting through a spin-dependent force,
as in the extended, attractive Hubbard model. A mean field
calculation in the Hartree-Fock Bogoliubov approximation suggests
that the superfluid ground state generated by strong nucleon
pairing undergoes a second-order phase transition to a normal
state as the temperature increases. The calculation is shown to
lead to a promising description of the thermal properties of
low-density neutron matter.  A possibility of a density wave phase
is also examined.

\end{abstract}
\maketitle
\section{Introduction}
Nuclear excitations are complicated dynamical phenomena, depending
on the detailed structure of the individual nucleus, and must be
examined based on the specific structure of the nucleus, such as
whether it is closed or open shell. As the excitation energy gets
higher, however, the excitations depend less on specific nuclear
structure and start to exhibit more common features among (heavy)
nuclei. These features are expected to be reasonably well
represented by excitations of nuclear matter.  Furthermore,
dynamics of supernovae and neutron stars, which have been of much
astronomical interest, are expected to be better understood
through the study of excitations of neutron matter
\cite{dean,arpns}. The gross features of the thermal properties of
nucleon matter have been examined by means of statistical models
\cite{stat} and lattice gas models \cite{latgas}. More realistic
descriptions of the thermal properties have been provided through
applications of various approaches in the nuclear many-body
theories \cite{TTT,panda-t,kyoto-t,other-t,jkuo,gapref}.
Applications of the traditional nuclear many-body theories
regarding nucleon matter at zero temperature have been extensive
\cite{arpns,panda,kyoto,fantoni}, and provide the most reliable
information on the properties of nucleon matter at low
temperatures.

Previously, one of us (R.~S.) has collaborated on a Monte Carlo
calculation of nuclear matter on a lattice \cite{MKSV}, which
provides a new framework for studying the thermal properties of
nucleon matter. Though the computational space was small and the
nucleon-nucleon interaction was simple, the calculation has proven
to be of much promise, demonstrating the occurrence of a phase
transition around 15 MeV using the parameters adjusted to
reproduce the saturation properties. A similar calculation of
nucleon matter of a single species has also been initiated in the
same work, using the same form of the Hamiltonian. Though a phase
transition appeared to take place at a few MeV, the evidence for
it was not quite solid owing to statistical fluctuations, which
are enhanced at low temperature (a sign problem).  The phase
transitions may correspond to those expected through paired
nucleons (Cooper pairs) in nucleon matter \cite{dean,arpns}.

In order to gain a better understanding of the possible, latter
phase transition, we apply in this work the analytic means of a
mean field approach to the problem in the same lattice
formulation. From this work, we do not expect to be able to draw
precise quantitative conclusions, but rather we will try to learn
the nature of the phase transition at a semi-quantitative level.
For this purpose, we take the thermodynamical (infinite volume)
limit for numerical results, so as to obtain a clear signal of the
phase transition. The mean field results in this work will also
serve as a reference for the more extensive Monte Carlo
calculation that we are currently carrying out.

Our Hamiltonian for single-species nucleon matter turns out to be
an extended, attractive Hubbard model, which has been studied as a
simple model of high-temperature superconductivity \cite{RMP:MRR}.
The mean field calculation shows that the low-temperature,
low-density state is a superfluid state and undergoes a continuous
(second-order) phase transition to the normal state as the
temperature and/or density increases.  As the Hamiltonian is not
yet fully realistic and the values of the interaction parameters
are uncertain, our results are not quite comparable to those for
neutron matter, except perhaps at a very low density. But we
demonstrate that the approach is promising for the study of
low-density neutron matter.

Furthermore, we find that a density-wave state coexists with the
superfluid state, suggesting that the state of neutron matter may
be more complicated than the simple description of a superfluid
state as it is often characterized.

The outline of this work is as follows: After the introduction in
Sec.~I, the Hamiltonian and its discretized form in the
coordinate space are presented and are identified as an extended
Hubbard model in Sec.~II. The Hartree-Fock Bogoliubov
approximation is applied and the Hamiltonian is diagonalized in
Sec.~III. Thermodynamical properties numerically calculated are
shown in Sec.~IV, an attempt to apply our calculation to the
problem of low-density neutron matter is discussed in Sec.~V, 
and the possibility of a density wave phase is examined in Sec.~VI.  
Discussions and conclusion are presented in Sec.~VII.

\section{Simplified Hamiltonian and Extended Hubbard Model}
\hspace*{\parindent}The Hamiltonian consists of the kinetic and
potential terms ${\hat K}$ and ${\hat V}$, respectively:
\begin{eqnarray}
  {\hat H} &=& {\hat K} + {\hat V} \; \nonumber \\
    &=& - \frac{\hbar^2}{2m_N} \sum_{\sigma\tau}
    \int {\rm d}{\bf r} \;
    {\hat \psi}^\dagger_{\sigma\tau}({\bf r})
    {\nabla}^2 {\hat \psi}_{\sigma\tau}({\bf r}) \nonumber \\
  & & + \frac{1}{2} \sum_{\stackrel{\sigma \sigma^\prime}{\tau \tau^\prime}}
    \int {\rm d}{\bf r} \; \int {\rm d}{\bf r}^\prime \;
    {\hat \psi}^\dagger_{\sigma \tau}({\bf r})
    {\hat \psi}^\dagger_{\sigma^\prime \tau^\prime}({\bf r}^\prime)
    V({\bf r} - {\bf r}^\prime)
    {\hat \psi}_{\sigma^\prime \tau^\prime}({\bf r}^\prime)
    {\hat \psi}_{\sigma \tau}({\bf r}) \; ,
  \label{Hcr}
\end{eqnarray}
where $m_N$ is the nucleon mass, and $\sigma = \pm 1/2$ and $\tau
= \pm 1/2$ are the spin ($\uparrow$ or $\downarrow$) and the
isospin ($p$ or $n$), respectively. ${\hat \psi}_{\sigma
\tau}^\dagger({\bf r})$ and ${\hat \psi}_{\sigma \tau}({\bf r})$
are the creation and annihilation operators of the nucleon, with
the spin $\sigma$ and isospin $\tau$ at the position ${\bf r}$. As
in the previous Monte Carlo lattice calculation \cite{MKSV}, we
include only the central and spin-exchange interactions, $V_c$ and
$V_\sigma$, respectively:
\begin{equation}
  V({\bf r}-{\bf r}^\prime)
  = V_c({\bf r}-{\bf r}^\prime)
  + V_\sigma({\bf r}-{\bf r}^\prime)
     \mbox{\boldmath $\sigma$} \cdot \mbox{\boldmath $\sigma$}^\prime \; .
  \label{Vcs}
\end{equation}
$V_c$ and $V_\sigma$ are taken to consist of on-site and
next-neighbor interactions,
\begin{equation}
  \begin{array}{l}
  \displaystyle{
  V_c\left({\bf r}-{\bf r}^\prime\right)
  = V^{(0)}_c \delta \left({\bf r}-{\bf r}^\prime\right)
  + V^{(2)}_c \left[\nabla^2_{\bf r} \delta
  \left({\bf r}-{\bf r}^\prime\right)\right]
  } \\
  \displaystyle{
  V_\sigma\left({\bf r}-{\bf r}^\prime\right)
  = V^{(0)}_\sigma \delta \left({\bf r}-{\bf r}^\prime\right)
  + V^{(2)}_\sigma \left[\nabla^2_{\bf r} \delta
  \left({\bf r}-{\bf r}^\prime\right)\right] \; ,
  }
  \end{array}
  \label{Vs}
\end{equation}
where $V^{(2)}$ terms can be written explicitly exhibiting their
hermiticity.

As physics of the lattice description is more apparent in the
coordinate space, we consider the discretized coordinate with an
inter-nucleon spacing $a$ in the cubic lattice with the torus
boundary conditions. We thus focus our interest on the physics of
the spatial separation greater than $a$ in each direction, or of
the momentum component roughly between $\pi/a$ and $-\pi/a$, by
eliminating (or integrating out) the physics of the shorter
distance. The discretization corresponds to
\begin{eqnarray}
  {\bf r}            &\rightarrow& a {\bf n}_i  \;, \nonumber \\
  \int {\rm d}{\bf r} &\rightarrow& a^3 \sum_{i} \;, \nonumber \\
  \sum_{\sigma \tau} \int d {\bf r} \;
  {\hat \psi}_{\sigma \tau}^\dagger({\bf r})
  {\hat \psi}_{\sigma \tau}({\bf r})
  &\rightarrow&
  \sum_{i \sigma \tau}
  {\hat c}_{i \sigma \tau}^\dagger
  {\hat c}_{i \sigma \tau} \; , \nonumber
  \label{cubic}
\end{eqnarray}
where $i$ denotes a lattice site specified by ${\bf n}_i$ with its
component ranging $[-aN/2,aN/2]$. Here, $N$ is the number of sites
in each spatial direction. Note that the creation and annihilation
operators, ${\hat c}_{i \sigma \tau}^\dagger$ and ${\hat c}_{i
\sigma \tau}$, have no dimension as defined.  We also apply the
identity
\begin{equation}
\sum_i\sigma^{(i)}_{\alpha\beta}\sigma^{(i)}_{\gamma\delta}
 = 2\delta_{\alpha\delta}\delta_{\beta\gamma}
   -\delta_{\alpha\beta}\delta_{\gamma\delta} \;, \nonumber
\end{equation}
where $i$ denotes the spatial components $(x,y,z)$, and the Greek
indices denote the components of the Pauli spin matrix, 1 or 2.

In this work, we study the simplified case of nucleon matter
consisting of a single nucleon species, such as neutron matter.
The Hamiltonian Eq.~(\ref{Hcr}) is then expressed in a spatially
discretized form:
\begin{eqnarray}
  {\hat H} &=& - t \sum_{\langle i,j \rangle \sigma}
  {\hat c}_{i \sigma}^\dagger {\hat c}_{j \sigma}
      + 6t\sum_{i\sigma}
      {\hat c}_{i\sigma}^\dagger {\hat c}_{i\sigma}
   + \tilde{U} \sum_{i}
  {\hat c}_{i \uparrow}^\dagger
  {\hat c}_{i \downarrow}^\dagger
  {\hat c}_{i \downarrow}
  {\hat c}_{i \uparrow} \nonumber \\
  & & + \; V_1
      \sum_{\langle i,j \rangle \sigma}
      {\hat c}_{i\sigma}^\dagger
      {\hat c}_{j\sigma}^\dagger
      {\hat c}_{j\sigma}
      {\hat c}_{i\sigma}
  + V_2
      \sum_{\langle i,j \rangle \sigma}
      {\hat c}_{i\sigma}^\dagger
      {\hat c}_{j-\sigma}^\dagger
      {\hat c}_{j-\sigma}
      {\hat c}_{i\sigma}
   + V_3
      \sum_{\langle i,j \rangle \sigma}
      {\hat c}_{i\sigma}^\dagger
      {\hat c}_{j-\sigma}^\dagger
      {\hat c}_{j\sigma}
      {\hat c}_{i-\sigma} \; ,
\label{HUB0}
\end{eqnarray}
where $\langle i,j \rangle$ denotes the pairs of next-neighbor
sites, and $t$ is the hopping (kinetic energy) parameter defined
as
\begin{equation}
  t = \frac{\hbar^2}{2m_N a^2} \; .
  \label{t}
\end{equation}
Here, the potential parameters $\tilde{U}$ and $V$'s are expressed
in terms of linear combinations of $V^{(0)}$'s and $V^{(2)}$'s.
The Hamiltonian Eq.~(\ref{HUB0}) is now in the form of an extended
Hubbard model, which is the Hubbard model with the on-site
spin-pairing interaction of the $\tilde{U}$ term, modified by the
next-neighbor interaction of the $V$ terms.  As the $\tilde{U}$
value will be taken to be negative, our model is an extended,
attractive Hubbard model.  The repulsive Hubbard model has been
well studied in condensed matter physics as a model of strongly
correlated electron systems \cite{Fradkin}, but the attractive
model is generally less studied.  In recent years, however, the
extended, attractive Hubbard model has drawn much attention as the
model describing the essential features of high-temperature
superconductivity \cite{RMP:MRR}.  Note that the extended,
attractive (negative-$U$) Hubbard model used in condensed matter
physics, however, is usually of the simpler form shown below and
has no $6t$ term as a part of the kinetic energy \cite{tara}.

When the spin-dependent next-neighbor interaction is taken to be
small and negligible,
\begin{equation}
    V_\sigma^{(2)} = 0 \;,
\end{equation}
the Hamiltonian is simplified:
\begin{equation}
  {\hat H} = - t \sum_{\langle i,j \rangle \sigma}
  {\hat c}_{i \sigma}^\dagger {\hat c}_{j \sigma}
      + 6t\sum_{i\sigma}
      {\hat c}_{i\sigma}^\dagger {\hat c}_{i\sigma}
  + U \sum_{i}
  {\hat c}_{i \uparrow}^\dagger
  {\hat c}_{i \downarrow}^\dagger
  {\hat c}_{i \downarrow}
  {\hat c}_{i \uparrow}
  + V \sum_{\langle i,j \rangle \sigma \sigma^\prime}
  {\hat c}_{i \sigma}^\dagger {\hat c}_{i \sigma}
  {\hat c}_{j \sigma^\prime}^\dagger
  {\hat c}_{j \sigma^\prime} \; ,
  \label{HhubbardN}
\end{equation}
where
\begin{eqnarray}
  U &=& \frac{1}{a^3}\left( V^{(0)}_{c} - 6 \frac{V^{(2)}_{c}}{a^{2}}
                         - 3 V^{(0)}_{\sigma} \right)  \; , \nonumber \\
  V &=& \frac{1}{2a^5}V^{(2)}_c \; .
\end{eqnarray}

Apart from the lattice spacing $a$, the Hamiltonian of Eq.~
(\ref{HhubbardN}) now describes dynamics with two parameters. The
Hamiltonian Eq.~(\ref{HUB0}) [and thus also Eq.~(\ref{HhubbardN})]
possesses an underlying particle-hole symmetry, which affects
thermodynamical properties as discussed in Sec.~VI.   The
symmetry is not an explicit property in our original Hamiltonian,
Eqs.~(1)--(3). We elaborate on the symmetry in Appendix A.

\section{Hartree-Fock Bogoliubov Approximation and Gap Equations}
\label{HFB}
We now apply the mean field method in the Hartree-Fock Bogoliubov
approximation \cite{HFB}. Here, we expect effects of the $U$ term
to dominate the thermal properties of the single species matter as
in the standard BCS description \cite{schrieffer, Leg, Noz}, but
we also wish to treat their single-particle aspects in the
Hartree-Fock approximation on the same footing.  Since the method
is well known, we limit the description of the formalism to the
key steps that are specifically relevant to our calculation.

The nature of the mean field approximation is apparent in the
spatial representation.  Through the application of the Wick
theorem, the decoupling scheme for the $U$ term is
\begin{displaymath}
  {\hat c}_{i\uparrow}^\dagger {\hat c}_{i\downarrow}^\dagger
  {\hat c}_{i\downarrow} {\hat c}_{i\uparrow}
  \simeq
  - \Delta_{i}^\ast {\hat c}_{i\downarrow} {\hat c}_{i\uparrow}
  - \Delta_{i}{\hat c}_{i\uparrow}^\dagger {\hat c}_{i\downarrow}^\dagger
  - \left| \Delta_{i} \right|^2
  + \; n_{i\uparrow} {\hat c}_{i\downarrow}^\dagger
  {\hat c}_{i\downarrow}
  + n_{i\downarrow} {\hat c}_{i\uparrow}^\dagger {\hat c}_{i\uparrow}
  - n_{i\uparrow} n_{i\downarrow}
  \; ,
  \label{dcp}
\end{displaymath}
where
\begin{eqnarray*}
  &&\Delta_{i}
  \equiv \langle {\hat c}_{i\uparrow}
                 {\hat c}_{i\downarrow} \rangle
  = - \langle {\hat c}_{i\downarrow}
              {\hat c}_{i\uparrow} \rangle \; , \\
  &&n_{i\uparrow}
  \equiv \langle {\hat c}_{i\uparrow}^\dagger
                   {\hat c}_{i\uparrow} \rangle
  \;\; {\rm and} \;\;
  n_{i\downarrow}
  \equiv \langle {\hat c}_{i\downarrow}^\dagger
                   {\hat c}_{i\downarrow} \rangle \; .
\end{eqnarray*}
Here, $\langle \;\cdots \; \rangle$ denotes the expectation value
in the BCS-like ground state, which is to be determined in a
self-consistent way. The order parameters $\Delta_i$'s are related
to the local density of the condensate of nucleon pairs, while
$n_i$'s are the average number of nucleons. Note that the
inclusion of $n_i$ as a variational parameter distinguishes the
present treatment from the standard BCS \cite{HFB}. For the $V$
terms, we have
\begin{displaymath}
  {\hat c}_{i \sigma}^\dagger {\hat c}_{i \sigma}
  {\hat c}_{j \sigma^\prime}^\dagger
  {\hat c}_{j \sigma^\prime}
  \simeq
  n_{i\sigma}
  {\hat c}_{j\sigma^\prime}^\dagger {\hat c}_{j\sigma^\prime}
  + n_{j\sigma^\prime}
  {\hat c}_{i\sigma}^\dagger {\hat c}_{i\sigma}
  -  n_{i\sigma} n_{j\sigma^\prime} \; .
\end{displaymath}

In the following, $\Delta_i$ and $n_i$'s will be assumed to be
independent of the site $i$, or global, and will be formally
treated as the variational parameters:
\begin{eqnarray*}
     &&\Delta \simeq 2U \Delta_i = 2U \Delta_i^\ast \;, \\
     &&n \simeq 2 n_{i\uparrow} = 2 n_{i\downarrow} \; .
\label{simapprox}
\end{eqnarray*}
Our $\Delta$ is defined to be the gap energy itself, with the
dimension of energy in the unit of $2U$, and carries an extra
factor of 2 in comparison to the often-used $\Delta$. We also note
that we ignore the $V$-term contribution to the hopping term, as
they merely change somewhat the strength of the hopping term and
of the constant part of the energy, without affecting the physics
of the phase transition. The mean field approximation then yields
in a cubic lattice with six neighboring sites
\begin{eqnarray}
  {\hat H}
  &=& - t \sum_{\langle i,j \rangle \sigma}
           {\hat c}_{i \sigma}^\dagger {\hat c}_{j \sigma}
   + \left(6t+\frac{n}{2}{\bar U}\right) \sum_{i\sigma}
      {\hat c}_{i\sigma}^\dagger {\hat c}_{i\sigma} \nonumber \\
  & & - \frac{\Delta}{2} \sum_{i} \left(
  {\hat c}_{i \uparrow}^\dagger
  {\hat c}_{i \downarrow}^\dagger +
  {\hat c}_{i \downarrow} {\hat c}_{i \uparrow}
  \right)
  - \frac{N^3}{4} \left( \frac{\Delta^2}{U}
  + n^2{\bar U} \right) \; ,
  \label{freeC}
\end{eqnarray}
where ${\bar U} \equiv U + 24 V$.

For convenience, we will carry out the mean field calculations in
the momentum space. The momentum representation is introduced by
discretizing the momentum as ${\bf p} \rightarrow  2\pi{\bf
k}/(Na)$, with each component of ${\bf k}$ being an integer,
ranging $[ -N/2,N/2]$. Note that we now have ${\bf r}\cdot{\bf p}
\rightarrow 2\pi{\bf n}_i \cdot {\bf k}/N$. The coordinate and
momentum representations of the operator $\hat c$ are related
through the Fourier transformations
\begin{equation}
  \begin{array}{l}
  \displaystyle{
  {\hat c}_{j\sigma} = \frac{1}{\sqrt {N^3}} \sum_{\bf k}
  e^{-2i{\pi}{\bf k} \cdot {\bf n}_j/N}
  {\hat c}_{{\bf k} \sigma} \; ,
  } \\
  \displaystyle{
  {\hat c}_{{\bf k} \sigma} = \frac{1}{\sqrt {N^3}} \sum_j
  e^{2i{\pi}{\bf k} \cdot {\bf n}_j/N}
  {\hat c}_{j \sigma} \;,
  }
  \end{array}
  \nonumber
\end{equation}
and similarly for the nucleon creation operator, $\hat c^\dagger$.
Note that the discretized orthonormality relation is
\begin{equation}
  \sum_{j} \exp \left( 2i \pi{\bf k} \cdot {\bf n}_j /N \right)
  = N^3 \delta_{{\bf k},{\bf 0}} \; .
  \label{deltafunc}
\end{equation}

With the chemical potential $\mu$, ${\hat H} -\mu N^3 {\hat n}$ is
in the momentum space
\begin{eqnarray}
  \hat{H} -\mu N^3 {\hat n}
  &=& \sum_{\bf k}
  \left( {\hat c}_{{\bf {k}}\uparrow}^\dagger
  \;\;\; {\hat c}_{{\bf {-k}}\downarrow} \right)
  \left(
  \begin{array}{cc}
  \epsilon_{\bf k}-{\bar \mu} &  - \frac{1}{2}\Delta              \\
     - \frac{1}{2}\Delta      &  -( \epsilon_{\bf k}-{\bar \mu})
  \end{array}
  \right)
  \left(
  \begin{array}{c}
  {\hat c}_{{\bf {k}}\uparrow} \\
  {\hat c}_{{\bf {-k}}\downarrow}^\dagger
  \end{array}
  \right) \nonumber\\
  & &  - \frac{N^3}{4} \left( \frac{\Delta^2}{U}
  + n^2{\bar U} \right) \; ,
  \label{gscfH}
\end{eqnarray}
where
\begin{equation}
{\bar \mu} =  \mu -\left(6t+n{\bar U}/2 \right)\;, \label{mubar}
\end{equation}
and
\begin{equation}
  \epsilon_{\bf k} = - t \sum_{\bf e} \exp \left( 2i\pi
  {\bf k} \cdot {\bf e }/N \right)
  = -2t \sum_{j=x,y,z} \cos \left( 2\pi k_j/N \right)
  \label{epsilon}
\end{equation}
is a part of the kinetic energy of a quasi-particle expressed in
terms of the unit vector ${\bf e}$ showing a next-neighbor site.
Note that $\epsilon_{\bf {-k}} = \epsilon_{\bf k}$ and $\sum_{\bf
k} \epsilon_{\bf k} = 0$. We see that in the Hamiltonian, $V$
appears only as ${\bar U} = U + 24V$, and merely shifts the
chemical potential and the total energy: It does not actively
participate in the generation of the phase transition.

As is well known, the spin density $s$,
\begin{equation}
s =\langle \hat{s} \rangle
  \equiv  \frac{1}{2N^3} \sum_{i}
  \langle{\hat c}_{i \uparrow}^\dagger
  {\hat c}_{i \uparrow}
  -{\hat c}_{i \downarrow}^\dagger
  {\hat c}_{i \downarrow}\rangle
  = \frac{1}{2N^3} \sum_{\bf k}
  \langle{\hat c}_{{\bf k}\uparrow}^\dagger
  {\hat c}_{{\bf k}\uparrow}
  -{\hat c}_{{\bf -k}\downarrow}^\dagger
  {\hat c}_{{\bf -k}\downarrow}\rangle \; ,
  \label{sdef}
\end{equation}
is conserved in the mean field approach, while $n$,
\begin{equation}
 n =\langle \hat{n} \rangle
   \equiv \frac{1}{N^3}\sum_{i\sigma} \langle
      {\hat c}_{i\sigma}^\dagger {\hat c}_{i\sigma}\rangle
   =\frac{1}{N^3}\sum_{{\bf k}\sigma} \langle
      {\hat c}_{{\bf k}\sigma}^\dagger {\hat c}_{{\bf k}\sigma}\rangle 
\label{ndef}
\end{equation}
(and thus the total nucleon number), is not, as the spin number
operator $\hat{s}$ commutes with $\hat{H}$ of Eq.~(\ref{freeC}),
but the number density operator $\hat{n}$ does not. We remedy this
problem by the standard method of introducing a Lagrange
multiplier corresponding to the chemical potential $\mu$, by
subtracting a term $\mu N^3 \hat{n} = \mu\sum_{i\sigma}{\hat
c}_{i\sigma}^\dagger {\hat c}_{i\sigma}$ from the $\hat{H}$. $\mu$
will be adjusted to achieve the desired value of the conjugate
parameter, $n$. The formalism is thus essentially the canonical
ensemble method.

We now diagonalize ${\hat H} -\mu N^3 {\hat n}$ by the use of the
Bogoliubov-Valatin transformation,
\begin{equation}
  \begin{array}{l}
  \displaystyle{
  {\hat \beta}_{{\bf k}+}^\dagger
  = u_{\bf k} {\hat c}_{{\bf k}\uparrow}^\dagger
  - v_{\bf k} {\hat c}_{{\bf -k}\downarrow} \;, 
  } \\
  \displaystyle{
  {\hat \beta}_{{\bf -k}-}^ \dagger
  = u_{\bf k} {\hat c}_{{\bf -k}\downarrow}^\dagger
  + v_{\bf k} {\hat c}_{{\bf k}\uparrow} \;,
  }
  \end{array}
  \label{betas}
\end{equation}
where $u_{\bf k}$ and $v_{\bf k}$ are taken to be real and are
given by
\begin{displaymath}
  u_{\bf k}^2 = \frac{1}{2} \left( 1 +
  \frac{\epsilon_{\bf k} - {\bar \mu}}{E_{\bf k}} \right)
  \;\; {\rm and} \;\;
  v_{\bf k}^2 = \frac{1}{2} \left( 1 -
  \frac{\epsilon_{\bf k} - {\bar \mu}}{E_{\bf k}} \right) \; ,
\end{displaymath}
satisfying $u_{\bf k}^2 + v_{\bf k}^2 = 1$. ${\hat H} -\mu N^3
{\hat n}$ is now expressed as that of a system of free
quasi-particles:
\begin{equation}
{\hat H} -\mu N^3 {\hat n} = \sum_{{\bf k}\lambda=\pm} E_{\bf k}
  {\hat \beta}_{{\bf k}\lambda}^\dagger {\hat \beta}_{{\bf k}\lambda}
  + N^3 (E_{\rm GS} - \mu n) \; ,
  \label{Hcan}
\end{equation}
with the energy of a quasi-particle
\begin{equation}
  E_{\bf k} = \sqrt{(\epsilon_{\bf k}-{\bar \mu})^2
  + {\Delta}^2/4} \;
  \label{Ek}
\end{equation}
and the ground-state energy of the system
\begin{equation}
  E_{\rm GS} = - \frac{1}{4} \left( \frac{\Delta^2}{U}
               + n^2{\bar U}\right)
               - \frac{1}{N^3} \sum_{\bf k} E_{\bf k} - {\bar \mu}
               + \mu n\;.
  \label{Egs}
\end{equation}
Equation (\ref{betas}) shows that $\hat{\beta}_{{\bf k}\lambda}$'s
obey anti-commutation relations and that the quasi-particles are
fermions. Furthermore, Eq.~(\ref{Hcan}) implies that they form a
system of free fermions.  As a consequence of the thermal average,
the internal energy is then given by
\begin{equation}
  E \equiv
  \langle \hat{H} \rangle
  = \sum_{{\bf k}\lambda} E_{\bf k} n_{{\bf k}\lambda}
  + N^3 E_{\rm GS} \; ,
  \label{Have}
\end{equation}
with the energy per lattice site being $E/N^3$.  Here, $n_{{\bf
k}\lambda}$ is the momentum distribution of the quasi-particles,
$n_{{\bf k}\lambda} \equiv \langle {\hat \beta}_{{\bf
k}\lambda}^\dagger
  {\hat \beta}_{{\bf k}\lambda} \rangle$,
and is determined by the requirement that the free energy $F$
introduced below is minimized by a variation of $n_{{\bf
k}\lambda}$, $\delta F/\delta n_{\bf k \lambda} = 0$ for $\lambda
= \pm$.  We obtain
\begin{equation}
  n_{\bf k} \equiv n_{{\bf k}+} = n_{{\bf k}-}
   = \left[ \exp\left( E_{\bf k}/T \right)  + 1 \right]^{-1} \;,
  \label{nval}
\end{equation}
which have the limiting values, $n_{\bf k} \rightarrow 0$ and
$\rightarrow 1/2$ as $T \rightarrow 0^+$ and $\rightarrow \infty$,
respectively. Note that throughout this work, we denote the
temperature $T$ in the unit of the Boltzmann constant.

By combining with the entropy $S$, the (Helmholtz) free energy is
expressed as
\begin{eqnarray}
  F(T,a;\Delta,n) &\equiv& E - TS
\nonumber \\
  &=& \sum_{{\bf k}\lambda} E_{{\bf k}}n_{{\bf k}\lambda}
  + N^3 E_{\rm GS} \nonumber \\
  & & + T \sum_{{\bf k}\lambda}
  \left[n_{{\bf k}\lambda} \ln n_{{\bf k}\lambda}
  + \left( 1 - n_{{\bf k}\lambda} \right) \ln
  \left( 1 - n_{{\bf k}\lambda} \right) \right] \;,
  \label{F}
\end{eqnarray}
depending on $a$ through the ${\bf k}$ sum because the spatial
volume $(aN)^3$ depends on $a$ with $N$ fixed. $F$ is a function
of $\Delta$ and $n$.  $\Delta$ and $\mu$ are determined so as to
minimize $F$ for variations of $\Delta$ and $n$, while $T$ and $a$
are fixed. The conditions
\begin{displaymath}
  \frac{1}{N^3}\frac{\partial F}{\partial n} = \mu
  \; , \quad
  \frac{\partial F}{\partial \Delta} = 0
\end{displaymath}
provide the gap equations
\begin{equation}
  \begin{array}{l}
  \displaystyle{
  n - 1 = \frac{1}{N^3}\sum_{\bf k}\frac{\epsilon_{\bf k}-{\bar \mu}}
  {E_{\bf k}} \left( 2n_{\bf k}-1 \right) \; ,
  } \\
  \displaystyle{
  \Delta \left( 1 - \frac{U}{2N^3} \sum_{\bf k}
  \frac{2n_{\bf k}-1}{E_{\bf k}} \right) = 0
  }
  \end{array}
  \label{gapeq}
\end{equation}
from which $\mu$ and $\Delta$ are determined.

\section{Thermodynamical properties}
\label{Thermo_properties}
We now apply the formalism so far described, to compute various
thermodynamical quantities.  For clarifying our presentation, we
place some expressions of the thermodynamical variables in
Appendix B. All numerical results are calculated in the
thermodynamical limit $N \rightarrow \infty$. In the limit, the
summation over the discretized momentum space of each component of
$\bf k$ ranging in $[-N/2, N/2]$ is replaced by the integral over the
first Brillioun zone with each component of the momentum $\bf p$
ranging $[-\pi /a, \pi /a]$:
\begin{displaymath}
  \frac{1}{N^3} \sum_{\bf k} \rightarrow
  \left(\frac{a}{2\pi}\right)^3
  \int_{-\pi /a}^{\pi /a}\int_{-\pi /a}^{\pi /a}\int_{-\pi /a}^{\pi /a}
  d^3{\bf p} \; .
\end{displaymath}

\subsection{Potential parameters}
We apply the parameter values used in the previous Monte Carlo
lattice calculation for nuclear matter \cite{MKSV}:
\begin{eqnarray}
  \begin{array}{l}
    V_c^{(0)} = -181.5 \;\; {\rm MeV} \; {\rm fm}^3 \;, \\
    V_c^{(2)} =  37.8 \;\; {\rm MeV} \; {\rm fm}^5 \;, \\
    V_{\sigma}^{(0)} = -31.25 \;\; {\rm MeV} \; {\rm fm}^3 \;, \\
    V_\sigma^{(2)} = 0 \; ,
    \end{array}
\label{potpara}
\end{eqnarray}
with the lattice spacing $a = 1.842 \;\;{\rm fm}$. These parameter
values give
\begin{displaymath}
  \begin{array}{l}
     t =    6.11 \;\; {\rm MeV} \;, \\
     U = - 24.74 \;\; {\rm MeV} \;, \\
     V =    0.89 \;\;{\rm MeV} \; .
     \end{array}
\label{tUV}
\end{displaymath}
The parameter values of Eq.~(\ref{potpara}) were chosen in Ref.
\cite{MKSV} so as to reproduce the saturation density and energy
of nuclear matter on a finite $4 \!\times\! 4 \!\times\! 4$
lattice for the same Hamiltonian as ours, Eqs.~(\ref{Hcr})--(\ref{Vs}). 
Our Hamiltonian has no explicit $\tau$-dependent term,
and the parameter values effectively include the strong
neutron-proton interactions for nuclear matter. The use of the
parameters is thus not quite adequate as a realistic description
of the nucleon matter of single species, such as the neutron
matter. Furthermore, finite lattice volume effects make the
thermodynamical limit ($N \rightarrow \infty$) calculation
different from the finite volume calculation. For comparison
purposes with the previous and future Monte Carlo calculations,
however, we use the above parameter values except when the $U$
dependence of $\Delta$ is examined. Any conclusion that we could
draw from the numerical results in this section is then
qualitative.

\subsection{Gap parameter $\Delta$}
\begin{figure}[htbp]
\begin{center}
\includegraphics[width=100mm]{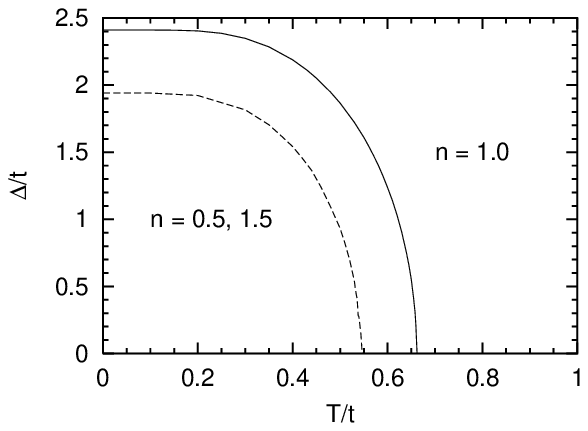}
\caption{The order parameter $\Delta$ as a function of temperature
$T$ in the unit of hopping parameter $t$ for the density $n = 0.5$
and 1.5 (long-dashed curve) and $n = 1.0$ (solid curve).}
\label{Fig:Delta_T}
{~~~}\\
{~~~}\\
{~~~}\\
{~~~}\\
\includegraphics[width=100mm]{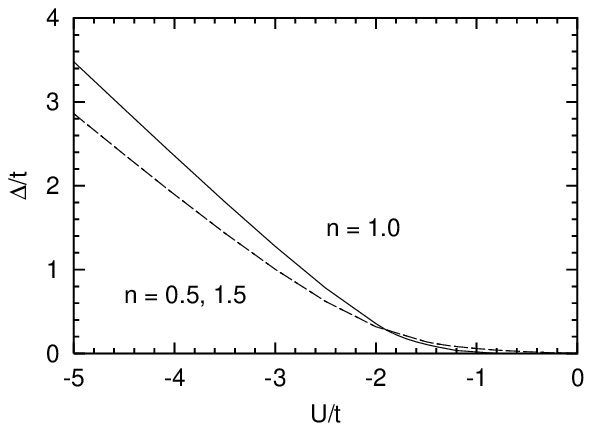}
\caption{The order parameter $\Delta$ as a function of the on-site
potential strength $U$ in the unit of hopping parameter $t$ at
zero temperature ($T= 0$) for the density $n = 0.5, 1.5$
(long-dash curve), $n = 1.0$ (solid curve).} \label{Fig:Delta_U}
\end{center}
\end{figure}

Equations (\ref{gapeq}) determine $\Delta$ and $\mu$. Figure
\ref{Fig:Delta_T} illustrates $\Delta$ as a function of the
temperature $T$ for $n =$ 0.5 (one-quarter filling), 1.0 (one-half
filling), and 1.5 (three-quarter filling). In the figure, we see
that $\Delta$ vanishes at $T = T_c$,
\begin{displaymath}
T_c = \left\{
      \begin{array}{rl}
      0.66 t \;\;{\rm or}\;\; 4.0 \; {\rm MeV} &
      \quad {\rm for} \;\; n = 1.0 \;, \\
      0.55 t \;\;{\rm or}\;\; 3.3 \; {\rm MeV} &
      \quad {\rm for} \;\; n = 0.5 \;\;{\rm and} \;\; 1.5 \; .
      \end{array} \right.
\end{displaymath}
The temperature dependence of $\Delta$ is the same for $n=0.5$ and
1.5.  This is a consequence of the symmetry with respect to $n=1$
and is discussed further in Sec.~IV D and Appendix A. Figure
\ref{Fig:Delta_T} also shows
\begin{displaymath}
\Delta (T = 0) \simeq (2 - 2.5) t \simeq 12 - 15 \; {\rm MeV} \;.
\end{displaymath}
The explicit values of $T_c$ and $\Delta(T = 0)$ depend sensitively
on the parameter values as discussed below, but $T_c$ and
$\Delta(T = 0)$ satisfy
\begin{displaymath}
  \Delta(T=0) \simeq 3.6 T_c \;,
\end{displaymath}
which is the well-known relation at the weak-coupling limit except
for the latter to have a slightly smaller coefficient 3.54
\cite{schrieffer}. In comparison to the weak limit, our
calculation thus somewhat underestimates $T_c$ in relation to
$\Delta(T=0)$.  Note that different mean field calculations have
been reported to yield the coefficient smaller than the weak-limit
value \cite{gapref} and also even much larger \cite{jkuo} than
ours.

Near $T_c$, the gap equations Eqs.~(\ref{gapeq}) yield
\begin{displaymath}
\Delta \propto \left\{
       \begin{array}{cl}
       (T-T_c)^{\beta} &
       \quad {\rm for}\;\; T < T_c \;, \\
       0 &
       \quad {\rm for}\;\; T > T_c \;,
       \end{array} \right.
\end{displaymath}
with $\beta \simeq 0.45$.  Note that the well-known mean field
value of the critical exponent $\beta$ in the simple BCS theory is
1/2 \cite{renorm}.

The physics of the phase transition depends on the strength of the
potential parameter $U$.  Figure \ref{Fig:Delta_U} illustrates how
sensitively the value of $\Delta$ depends on $U$.  We see that
$\Delta \rightarrow 0$ as $U \rightarrow 0$. As is well known,
$\Delta$ does not vanish for a finite $U$. In fact,
Eq.~(\ref{gapeq}) yields the well-known dependence of $\Delta$ on
$|U|$ for $U \rightarrow 0$:
\begin{displaymath}
   \Delta \rightarrow A e^{-B/|U|} \;,
\end{displaymath}
where $A$ and $B$ are constant and independent of $U$.

\subsection{Second-order (continuous) phase transition}
\begin{figure}[htbp]
\begin{center}
\includegraphics[width=100mm]{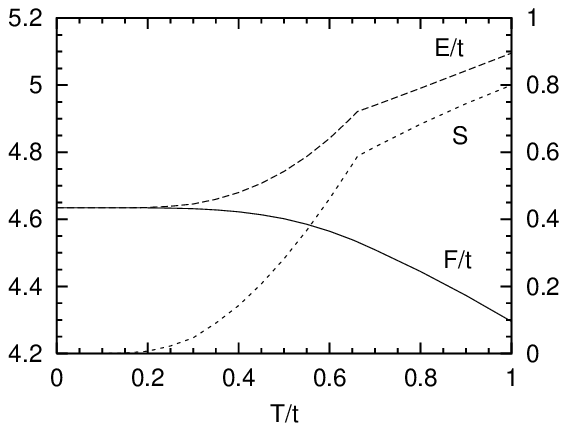}
\caption{The Helmholtz free energy $F$ (solid curve), the internal
energy $E$ (long-dashed curve), and the entropy of the system $S$
(short-dashed curve) as a function of temperature $T$ in the unit
of hopping parameter $t$ for the density $n = 1.0$. The left axis
refers to $F$ and $E$, and the right one is for $S$.}
\label{Fig:FES_T}
{~~~}\\
{~~~}\\
\includegraphics[width=100mm]{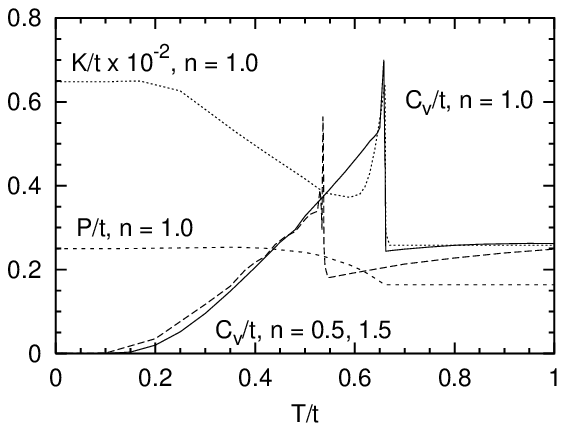}
\caption{The heat capacity $C_{v}$ for the density $n = 0.5$ and
1.5 (long-dashed curve) and $n = 1.0$ (solid curve), the pressure
$P$ for the density $n = 1.0$ (short-dashed curve), and the
incompressibility $K$ for the density $n = 1.0$ (dotted curve) as
a function of temperature $T$ in the unit of hopping parameter $t$
.} \label{Fig:CPK_T}
\end{center}
\end{figure}

The temperature dependence of $\Delta$ in Fig.~\ref{Fig:Delta_T}
is a well-known dependence of the order parameter for a
second-order phase transition. The variation of $\Delta$ as shown
in Fig.~\ref{Fig:Delta_T} implies that the phase transition takes
place from a superfluid state generated by spin pairing to the
normal state, as the temperature increases.  The features of the
second-order phase transition are clearly seen in the temperature
dependence of the thermodynamic quantities expressed in terms of
the temperature derivatives of the free energy $F$ in a successive
order. We consider the internal energy $E$, the entropy $S$, and
the heat capacity $C_v$,
\begin{equation}
  \begin{array}{l}
  \displaystyle{
  E = - T^2 \left[\frac{\partial (F/T)}{\partial T}\right]_{a,n} \;, 
  } \\
  \displaystyle{
  S = -\left[ \frac{\partial F}{\partial T}\right]_{a,n} \;, 
  } \\
  \displaystyle{
  C_v = -T \left[ \frac{\partial S}{\partial T}\right]_{a,n}
      = -T \left[ \frac{\partial^2 F}{\partial T^2}\right]_{a,n}
  \;.
  }
  \end{array}
\end{equation}
The temperature dependence of the quantities is calculated using
Eqs.~(\ref{Have}), (\ref{F}), (\ref{S}), and (\ref{C}), and is
illustrated in Figs.~\ref{Fig:FES_T} and \ref{Fig:CPK_T}. $E$ and
$S$ are continuous at the critical temperature $T_c$ as seen in
Fig.~\ref{Fig:FES_T}, while $C_v$ has a jump at $T_c$ as in
Fig.~\ref{Fig:CPK_T}. These behaviors demonstrate the generic
features of the second-order phase transition. The amount of the
discontinuity in $C_v$ at $T_c$, $\Delta C_v$, is relative to
$C_v$ of the normal phase,
\begin{displaymath}
  \Delta C_v / C_v ({\rm normal})
       \simeq \left\{
       \begin{array}{rl}
       1.74 &
       \quad {\rm for}\; n = 1.0 \;, \\
       1.43 &
       \quad {\rm for}\; n = 0.5 \;\;{\rm and}\;\; 1.5 \;,
       \end{array} \right.
\end{displaymath}
while the BCS mean field value is $12/7\zeta(3) \simeq 1.43$,
independent of $n$ \cite{schrieffer}. ($\zeta$ is the zeta
function.)

The thermal quantities involving volume derivatives form a set of
quantities similar to the temperature derivatives.  We consider
the pressure $P$ and the isothermal compressibility $\kappa_T$.
Here, following the common practice in nuclear physics, we examine
the incompressibility $K \equiv 9/(\kappa_T \rho)$, defined in
terms of the density $\rho = \mathcal{V}/(a N)^3 = n/ a^3$ with
the spatial volume $\mathcal{V}=(aN)^3$. A volume derivative is
then a derivative with respect to the lattice spacing $a$.  A
derivative with respect to $a$ requires, however, the knowledge of
$a$ dependence of $U$ and $V$, that is, their renormalization flow
when $a$ is varied. In this work, for simplicity, we assume that
their $a$ dependence is small, at least around the value of $a$ we
use. $P$ and $K$ are written as
\begin{eqnarray}
 P &=& -\left[ \frac{\partial F}{\partial \mathcal{V}}\right]_{T,a} \;, \\
 K &=& -9 \frac{\mathcal{V}}{\rho}\left[ \frac{\partial P}
          {\partial \mathcal{V}}\right]_{T,a}
     = -9\frac{\mathcal{V}}{\rho}\left[ \frac{\partial^2 F}
         {\partial\mathcal{V}^2}\right]_{T,a} \;.
\label{PK}
\end{eqnarray}
$P$ is calculated using Eq.~(\ref{P}) and $K$ is obtained
numerically from the temperature dependence of $P$. The
temperature dependence of $P$ and $K$ confirms the second-order
phase transition, as seen in  Fig.~\ref{Fig:CPK_T}.

There are many other quantities that describe thermodynamic
properties of the system described by our Hamiltonian, but they
are either related to the quantities already shown, or their
features depend strongly on the explicit form of the Hamiltonian.
We thus do not show them in this exploratory work.  For example,
the temperature dependence of double occupancy per site,
\begin{equation}
  D \equiv  \frac{1}{N^3}\sum_i
  \langle {\hat c}_{i\uparrow}^\dagger{\hat c}_{i\uparrow}
  {\hat c}_{i\downarrow}^\dagger{\hat c}_{i\downarrow}\rangle\;,
\end{equation}
provides the amount of the spin pairing that participates in the
phase transition.  But $D$, as well as the kinetic energy per
site, is related to $\Delta$ and $n$ in the mean field
approximation \cite{armen} and provides no new information, as
shown in Appendices A and B.

\subsection{Particle-hole symmetry and phase diagram}
\begin{figure}[htbp]
\begin{center}
\includegraphics[width=100mm]{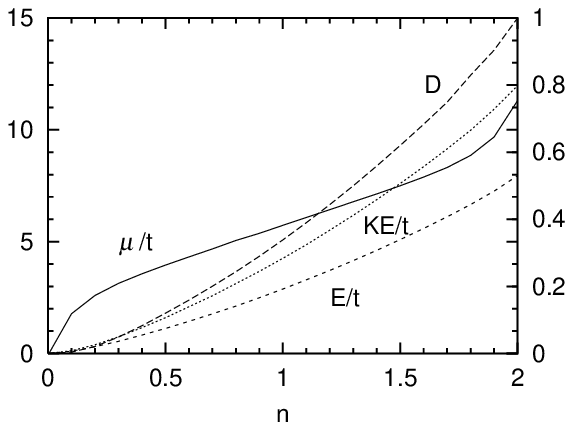}
\caption{The chemical potential $\mu$ (solid curve), the double
occupancy $D$ (long-dashed curve), the internal energy $E$
(short-dashed curve), and the kinetic energy $KE$ (dotted curve)
in the unit of hopping parameter $t$ as a function of the density
$n$ at zero temperature ($T = 0$). The left axis refers to $\mu$,
$E$, and $KE$, and the right one is for $D$.} \label{Fig:muDEKE_n}
{~~~}\\
{~~~}\\
Fig. 6
\includegraphics[width=100mm]{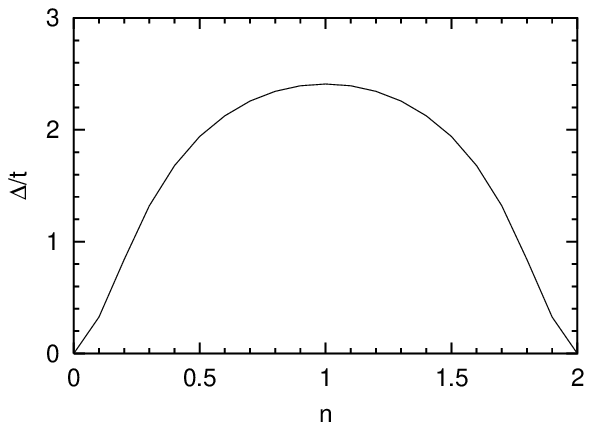}
\caption{$\Delta$ in the unit of hopping parameter $t$ as a
function of the density $n$ at zero temperature ($T = 0$).}
\label{Fig:Delta_n}
\end{center}
\end{figure}

For illustrative purposes, however, we show the density dependence
of $\mu$, $D$, $E$, and $KE$ at $T = 0$, in
Fig.~\ref{Fig:muDEKE_n}. 
As noted above, Fig.~\ref{Fig:Delta_T} suggests a symmetric
dependence of $\Delta$ on the density $n$, which is more clearly
illustrated in Fig.~\ref{Fig:Delta_n}. The symmetry is generated
as a consequence of the particle-hole symmetry, as seen from the
fact that Eq.~(\ref{gapeq}) is invariant under the particle-hole
conjugation though the symmetry is implicit in our Hamiltonian,
Eqs.~(\ref{HUB0}) and (\ref{HhubbardN}). There is a group of
Hamiltonians, in which the particle-hole symmetry is implicit, yet
yielding (in the mean field results) the energy spectrum of the
system with the explicit symmetry.  The Hamiltonians do yield
different behaviors of thermodynamic variables.
Figure\;\ref{Fig:muDEKE_n} is an example. We elaborate on the
issue of this symmetry in Appendix A.

Some thermodynamic variables are made to exhibit this symmetry
explicitly by modifying the Hamiltonian to possess the explicit
symmetry. Whether it is explicit or not, however, the
thermodynamic properties obtained from the Hamiltonian are
affected by the symmetry.  The symmetry is a consequence of our
computational method using the lattice configuration, and it is an
artifact.  In order to extract physically realistic results, we
should therefore stay away from the region of the symmetry and
should confine ourselves to a small value of $n$ by appropriately
adjusting the value of the lattice spacing $a$, so as to simulate
the desired density of the nucleon matter. We discuss this point
again in the following section, where we attempt to apply our
calculation to a case of low-density neutron matter.

\begin{figure}[htbp]
\begin{center}
\includegraphics[width=100mm]{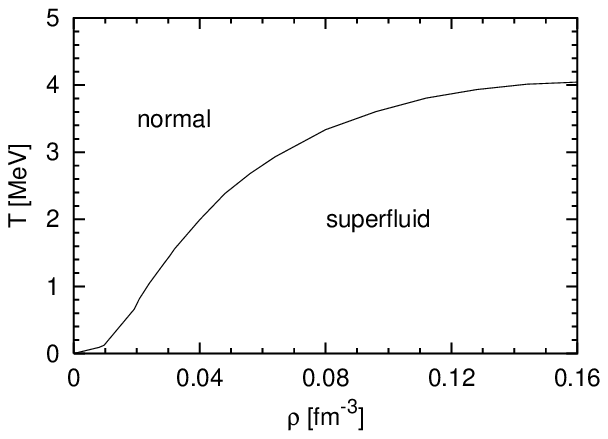}
\caption{$T$-$\rho$ phase diagram.}
\label{Fig:T_rho}
{~~~}\\
{~~~}\\
{~~~}\\
{~~~}\\
\includegraphics[width=100mm]{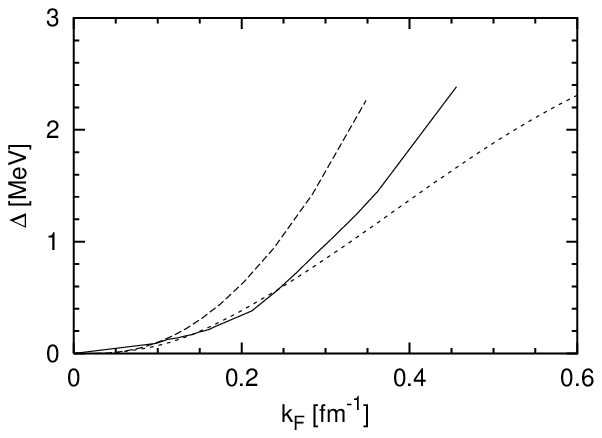}
\caption{The gap energy $\Delta$ as a function of the Fermi
momentum $k_F$ for $U = - 45.8$ MeV (solid curve) in comparison to
a simple BCS calculation (long-dashed curve) and a more elaborate
one \cite{gapref} (short-dashed curve).  The simple calculation is
taken from Ref. \cite{pb} [using Eq.~(8) with the neutron mass of
940 MeV and the neutron-neutron scattering length of -18.8 fm].
$\Delta$ in this figure is half the $\Delta$ defined in the text,
so that its definition agrees with the one used in \cite{gapref,pb}.} 
\label{Fig:Delta_kF_2}
\end{center}
\end{figure}

Combining the variations of the thermodynamic quantities, some of
which have been presented so far, we obtain the phase diagrams of
the present system described in the mean field theory.
Figure\;\ref{Fig:T_rho} shows the phase diagram in the region of
small densities where we expect the above-mentioned symmetry to be
generating less distortion.

\section{Low-density Neutron matter}
In the previous section, we have used the parameter values most
appropriate as a description of nuclear matter and have examined
the nature of the single species nucleon matter described by our
model.  In this section, we discuss whether our model could be
made a realistic description of neutron matter.

First, we have the question of whether our lattice would meet the
basic momentum requirement imposed by the lattice spacing.  A
lattice description can be made realistic when the lattice spacing
is less than the momentum scale of the system. By taking the Fermi
momentum of the neutron matter as an estimate of the momentum
scale, we have
\begin{equation}
\pi/a > p_F \;. \label{realistic}
\end{equation}
Our lattice spacing, $a = 1.842$ fm, yields the density of $\rho =
1/a^3 \simeq 0.160 \;\;{\rm fm}^{-3}$ for $n=1.0$ (the lattice
space being half full). The Fermi momentum corresponding to this
density is
\begin{equation}
p_F \simeq 1.68 \;\; {\rm fm}^{-1} \nonumber
 \label{fermi}
\end{equation}
using $\rho = {p_F}^3/3\pi^2$.  The value of $p_F$ is practically
the same as
\begin{equation}
\pi/a \simeq 1.71 \;\; {\rm fm}^{-1} \;, \nonumber
\end{equation} and
thus the lattice with the above lattice spacing is applicable to a
density much smaller than $\rho \simeq 0.160 \;\;{\rm fm}^{-3}$.
Note that the preceding discussion yields that the condition
\begin{equation}
         \pi/3 > n  \nonumber
\end{equation}
meets Eq.~(\ref{realistic}) independently of $a$.

Second, there is the question of whether our Hamiltonian is
appropriate for a realistic description of low-density neutron
matter.  The nucleon-nucleon ${}^1S_0$ phase shift is much greater
than the nucleon-nucleon phase shifts of other states below the
laboratory energy $E_{lab} \simeq$ 100 MeV. The nucleon momentum
in the center-of-mass coordinate system $p_{cm}$ corresponding to
this $E_{lab}$ is about 1.2 fm ${\rm}^{-1}$ through $E_{lab} = 4
(p_{cm}^2/2m_N)$, and is smaller than the above $\pi/a \equiv
p_{cutoff} \simeq$ 1.7 ${\rm fm}^{-1}$. We thus infer that our
Hamiltonian form of the $S$-wave should be reasonable for neutron
matter of a density less than 0.17 ${\rm fm}^{-3}$, which
corresponds to the Fermi momentum of 1.7 ${\rm fm}^{-1}$.

As to the parameter values in the Hamiltonian, it would be best to
determine them for our lattice size from experimental ${}^1S_0$
phase shifts by applying the method of effective field theory
\cite{MS,lstv}. Instead, as an exploratory study, we simply adjust
the $U$-parameter value so as to see whether our approach could
come close to other mean field calculations of low-density neutron
matter in the literature. Figure \ref{Fig:Delta_kF_2} illustrates
that we could obtain a somewhat reasonable density dependence of
$\Delta$ by increasing the magnitude of $U$.  We leave a more
serious determination of the parameters for our future work.

\section{Density wave phase}
Our analysis has so far been strictly based on the mean field
approximation applied to a spin-pairing phase at the same site. In
the approximation, the $V$ term merely shifts the effective
chemical potential and is inactive in generating the phase
transition.  As the $V$ term represents the pairing of the spin
densities at the adjacent sites, such a role may be a reasonable
one in this phase transistion.  Would the $V$ term ever play an
active role in generating a different phase transition? In this
section we briefly examine this possibility.

The most likely phase in which the $V$ term would play the major
role would be a density wave phase generated by a coupling of the
densities of the opposite spins at the adjacent sites.  We examine
how the $V$ term could generate such a phase transition, again in
the mean field approximation, and see whether the phase transition
would occur with our parameter values.

We follow the Hartree-Fock Bogoliubov approximation with the same
decoupling scheme as before.  For simplicity, however, we ignore
the superfluid phase. The density wave phase is introduced by
making a replacement,
\begin{eqnarray}
  \langle {\hat c}_{i\sigma}^\dagger {\hat c}_{i\sigma}\rangle
  &\rightarrow& n_{i\sigma}
  + (\delta/2)\cos(2\pi{\bf n}_i\cdot{\bf q} /N )
  \nonumber \\
  &\simeq& n/2
  + (\delta/2)\cos(2\pi{\bf n}_i\cdot{\bf q} /N )\;.
\end{eqnarray}
Here, the inhomogeneous order parameter for the density wave
depends on the amplitude $\delta$ and the wave number vector ${\bf
q}$. In the following, we examine the wave modes in which the
adjacent sites are the maximum and minimum of the amplitude in a
cubic lattice. That is, at least one component of $2\pi{\hat q}/N$
is $\pm\pi$. The number of the non-zero components, or the
dimension of the density wave $d$, provides a convenient parameter
$\eta$,
\begin{equation}
    \eta = 3 - 2d\;.
\end{equation}
As in the previous spin-pairing case, $n$ and $\delta$ are treated
as independent parameters, and $\mu$, the parameter conjugate to
$n$, is adjusted to the desired value of $n$.

The steps to the gap equations are similar to the case of the
superfluid phase discussed in Sec.~III, and are shown in
Appendix C.  By minimizing $F - \mu N^3 n$, we obtain
\begin{equation}
  \begin{array}{l}
  \displaystyle{
    n = \frac{1}{N^3}\sum_{{\bf k}\lambda}
        n_{{\bf k}\lambda} \;,
  } \\
  \displaystyle{
    \delta \left( 1 - \frac{ U + 8 \eta V}{N^3}
    \sum_{\bf k}
    \frac{n_{\bf k+}-n_{\bf k-}}{\xi_{\bf k}}\right) = 0
  } \\
  \end{array}
\label{gap2}
\end{equation}
by varying $n$ and $\delta$, respectively.  Equation (\ref{gap2})
determines $\delta$ and $\mu$. Furthermore, as $n_{\bf k+} <
n_{\bf k-}$ and
\begin{equation}
   U + 8\eta V < 0
\end{equation}
for our parameter values, Eq.~(\ref{gap2}) shows that the density
waves of all dimensions are expected to occur owing to the strong,
attractive $U$.

\section{Discussion and Conclusion}
The form of the nucleon-nucleon interaction, Eqs.~(1)--(3),
shows that the nucleon-nucleon interaction used in this work is of
the S states. As is well known, the realistic nucleon-nucleon
interaction is highly state-dependent. The relevant nucleon energy
of interest to us here is a few hundred MeV, corresponding to the
Fermi energy region of the nuclear matter density.  In this energy
region, the attractive neutron-neutron interaction is known to be
dominated by the ${}^3P_2$ interaction driven by the spin-orbit
force, coupled with the ${}^3F_2$ interaction associated with the
tensor force \cite{NN,TT}. Our Hamiltonian accommodates none of
these features of the interaction.  Our objective as noted in
Sec.~I is to understand the essential physics associated with
the thermal properties of nucleon matter, but our finding in this
work is limited in this sense and is perhaps most applicable to
low-density neutron matter.

There is a serious question of how good the mean field calculation
is in our case.  A mean field approximation ignores most features
of particle correlations.  As the particle correlations are the
vital ingredient of critical phenomena, the mean field
approximation is generally believed to be only of qualitative use,
and some times not even qualitative, as fluctuations could alter
the nature of the phase transition. The situation, however,
depends on the nature of the problem \cite{renorm}, as the
prominent success of the Ginzburg-Landau/BCS theory shows
\cite{superc}, especially at temperatures not too close to the
critical one.  Our problem is in three dimensions, close to the
usual upper critical dimension of four under the Ginsburg
criterion \cite{renorm}. We are hoping that our calculation, being
similar to the BCS theory, is not far off, but this remains to be
seen. This issue is under further investigation by incorporating a
renormalization approach, as has recently been done at zero
temperature \cite{schwenk}.

Our Hamiltonian has a form similar to that of the Skyrme
interaction \cite{skyrme} (though ours is a truncated form.) The
Skyrme interaction is one of the effective interactions that are
phenomenologically introduced to achieve quantitative agreement
with experiments, usually by the use of a mean field approximation
such as the Hartree-Fock calculations.  Though it is still not
quite realistic, our Hamiltonian has a justification in this
sense.  A lattice calculation such as the previous Monte Carlo
lattice calculation \cite{MKSV}, however, accounts for all the
complexity of the many-body interaction with no approximation
other than the numerical, in the lattice framework. The
nucleon-nucleon interaction used in it should not be then an
effective interaction like the Skyrme interaction, but an
interaction in free space.  The parameters $U$ and $V$ are
expected to be determined from scattering data through the use of
effective field theory by extending L{\"u}scher's formula
\cite{luescher,wash} for the large ${}^1S_0$ scattering length.
This issue is presently under investigation \cite{lstv}. Note that
the values of $U$ and $V$ will depend on the lattice spacing $a$.

In conclusion, our mean field calculation on a lattice with a
simple nucleon-nucleon interaction suggests a second-order phase
transition taking place at a low temperature in single species
nucleon matter described by a simple Hamiltonian. Thermodynamic
variables show their dependence on the temperature and density
variations as expected under the phase transition in the mean
field approach. The transition changes the phase of the matter
from a superfluid state due to nucleon pairing, to a normal state
as the temperature increases and the density decrease.  This
dependence is in qualitative agreement with the findings that have
been reported in the literature \cite{dean}.

\bigskip
\bigskip

\centerline{\bf ACKNOWLEDGMENTS}

\bigskip

This work is supported by the U.S.~Department of Energy under
Grant No.~DE-FG03-87ER40347 at CSUN, and the U.S.~National Science
Foundation under grants PHY0071856 and PHY0244899 at Caltech. T.
A. is supported under the 21st Century COE Program at the Tokyo
Institute of Technology, ``Nanometer-Scale Quantum Physics," by
the Ministry of Education, Culture, Sports, Science and Technology
in Japan. R.~S. acknowledges Satoshi Okamoto for patiently
explaining various aspects of condensed matter BCS theory. For his
warm hospitality, R. S. thanks Masahiko Iwasaki of RIKEN, where a
part of this work was done. T.~A. and R. S. acknowledge for its
hospitality the Institute for Nuclear Theory at the University of
Washington, where the last stage of this work was carried out.

\vfill\eject

\appendix

\section{The particle-hole symmetry}
The Hamiltonian of Eq.~(\ref{HhubbardN}) does not explicitly
exhibit particle-hole symmetry, but it can be modified to do so by
adding a single-particle Hamiltonian,
\begin{equation}
  \Delta{\hat H} = N^3 {\bar U}/4
                   - \left( 6t + {\bar U}/2 \right)
                   \sum_{i \sigma} \hat{n}_{i \sigma}\;,
  \label{Hadd}
\end{equation}
where $\hat{n}_{i \sigma} \equiv {\hat c}_{i\sigma}^\dagger {\hat
c}_{i\sigma}$. The new Hamiltonian,
\begin{eqnarray}
  {\hat H}^\prime
  &\equiv& {\hat H}+\Delta{\hat H} \nonumber \\
  &=& - t \sum_{\langle i,j \rangle \sigma}
  {\hat c}_{i \sigma}^\dagger {\hat c}_{j \sigma}
  + U \sum_{i}
  \left(\frac{1}{2} -{\hat n}_{i \uparrow}\right)
  \left(\frac{1}{2} -{\hat n}_{i \downarrow}\right)
  \nonumber \\
  & &+ V \sum_{\langle i,j \rangle \sigma \sigma^\prime}
  \left(\frac{1}{2}-{\hat n}_{i \sigma}\right)
  \left(\frac{1}{2}-{\hat n}_{j \sigma^\prime}\right)\;,
  \label{Heff}
\end{eqnarray}
is symmetric under the particle-hole conjugation (or with respect
to half filling, $n = 1$):
\begin{eqnarray}
  {\hat c}_{i \sigma}^\dagger &\leftrightarrow& {{\hat c}_{i
  \sigma}} \nonumber \\
  {\hat c}_{i \sigma} &\leftrightarrow& {{\hat c}_{i
  \sigma}^\dagger}\;,\nonumber
\end{eqnarray}
or $n_{i\sigma} \leftrightarrow  1-n_{i\sigma}$, which is $n
\leftrightarrow  2-n$ with $n \equiv \sum_{i\sigma} n_{i\sigma}$.
Under the conjugation, the Hamiltonian becomes that of holes, with
$t \rightarrow -t$.

When we repeat the BCS formulation in Sec.~III, after the
Bogoliubov transformation, we obtain the diagonalized form of the
new Hamiltonian, Eq.~(\ref{Heff}), as
\begin{equation}
{\hat H}^\prime = \sum_{{\bf k}\lambda} E_{\bf k}^{\prime}
  {\hat n}_{{\bf k}\lambda}^\prime
  + N^3 E_{\rm GS}^{\prime} \;
  \label{Hcansym}
\end{equation}
with the quasi-particle energy
\begin{equation}
  E_{\bf k}^\prime = \sqrt{(\epsilon_{\bf k}-{\bar \mu}^\prime)^2
  +{\Delta}^2/4} \;
  \label{Eksym}
\end{equation}
and
\begin{equation}
  n_{\bf k}^\prime \equiv n_{{\bf k}+}^\prime = n_{{\bf k}-}^\prime
               = \left[ \exp \left( E_{\bf k}^\prime/T \right)
                 + 1 \right]^{-1} \; .
  \label{nvalsym}
\end{equation}
Here, the ground-state energy is
\begin{equation}
  E_{\rm GS}^\prime = - \frac{1}{4}\left( \frac{\Delta^2}{U}
  + \frac{n^2 - 1}{4} {\bar U} \right)
  - \frac{1}{N^3} \sum_{\bf k} E_{\bf k}^\prime
  - {\bar \mu}^\prime + \mu n \;,
  \label{Egssym}
\end{equation}
with
\begin{equation}
  {\bar \mu}^\prime = \mu - (n-1){\bar U}/2 \;.
  \label{mubarprime}
\end{equation}

Equations (\ref{mubar}) and (\ref{mubarprime}) show that the $\mu$
value is shifted, and Eqs.~(\ref{Egs}) and (\ref{Egssym}) tell us
that the expression of the ground-state energy is altered.  The
kinetic energy is also changed:
\begin{equation}
  KE^\prime \equiv - \frac{t}{N^3}\sum_{\langle i, j \rangle, \sigma}
     \langle{\hat c}^\dagger_{i \sigma}{\hat c}_{j \sigma}\rangle
  = - \frac{1}{N^3} \sum_{\bf k} \frac{\left( \epsilon_{\bf k}
  - {\bar \mu}^\prime \right) \epsilon_{\bf k}}{E_{\bf k}^\prime}
  \;,
  \label{kineticsym}
\end{equation}
in comparison to Eq.~(\ref{kinetic}).  We also see that $E_{\rm
GS}^\prime$ and $KE^\prime$ for the new Hamiltonian explicitly
exhibit the particle-hole symmetry about half filling ($n=1$).

The double occupancy per site for the new Hamiltonian $D^\prime$,
\begin{equation}
  D^\prime \equiv  \frac{1}{N^3}\sum_i
  \langle {\hat c}_{i\uparrow}^\dagger{\hat c}_{i\uparrow}
  {\hat c}_{i\downarrow}^\dagger{\hat c}_{i\downarrow}\rangle
   = \frac{\Delta^2}{4U^2} + \frac{n^2}{4} \; ,
  \label{dsym}
\end{equation}
differs from that for the original Hamiltonian, Eq.~(\ref{d}).
Note that neither $D$ nor $D^\prime$ is symmetric about $n = 1$.

The gap equations for the new Hamiltonian become
\begin{equation}
  \begin{array}{l}
  \displaystyle{
  n - 1 = \frac{1}{N^3}\sum_{\bf k}\frac{\epsilon_{\bf k}-{\bar \mu}^\prime}
  {E_{\bf k}^\prime} \left( 2n_{\bf k}^\prime-1 \right) \;,
  } \\
  \displaystyle{
  \Delta \left( 1 - \frac{U}{2N^3} \sum_{\bf k}
  \frac{2n_{\bf k}^\prime-1}{E_{\bf k}^\prime} \right) = 0
  \; .}
  \end{array}
\end{equation}
These gap equations are the same as Eqs.~(\ref{gapeq}) except
${\bar \mu}$ is replaced by  ${\bar \mu}^\prime$.  As we solve the
gap equations for $\Delta$ and $\mu$ (or ${\bar \mu}$) for a fixed
$n$, the gap equations for the two Hamiltonians yield the same set
of $\Delta$ and ${\bar \mu}$, thus the same $T_c$.  Both
Hamiltonians thus provide the same excitation energy spectrum.

\bigskip
\bigskip
\bigskip
\section{Thermodynamic variables}
We list here the expressions of the thermodynamic variables, which
are used in Sec.~IV. The entropy is given 
in terms of $n_{{\bf k}\lambda}$ in Eq.~(\ref{nval}) by
\begin{equation}
 S = \sum_{{\bf k}\lambda}
  \left[n_{{\bf k}\lambda} \ln n_{{\bf k}\lambda}
  + \left( 1 - n_{{\bf k}\lambda} \right) \ln
  \left( 1 - n_{{\bf k}\lambda} \right) \right]  \; .
 \label{S}
\end{equation}
By the use of Eqs.~(\ref{Ek}), (\ref{Egs}), and (\ref{F}), the
heat capacity $C_v$ is expressed as
\begin{eqnarray}
 C_v &=& -T \left[ \frac{\partial^2 F}{\partial T^2}\right]_{a,n}
        \nonumber \\
 &=& \frac{1}{N^3} \sum_{\bf k} \frac{\Delta \Delta_T}{4 E_{\bf k}}
             \left( 2 n_{\bf k} - 1 \right)
        - \frac{2}{N^3} \sum_{\bf k} n_{\bf k}^2
    \left( \frac{\Delta \Delta_T}{4} - \frac{E_{\bf k}^2}{T^2} \right)
     e^{E_{\bf k}/T} - \frac{\Delta \Delta_T}{2U} \; .
  \label{C}
\end{eqnarray}
Here, $\Delta_T$ is
\begin{equation}
  \Delta_T \equiv \frac{\partial \Delta}{\partial T} \; ,
  \label{dt}
\end{equation}
and is numerically calculated from the solution of the gap
equations, Eq.~(\ref{gapeq}).

The pressure $P$ is written in terms of the space volume
$\mathcal{V} = (aN)^3$ as
\begin{eqnarray}
 P &=& -\left[ \frac{\partial F}{\partial \mathcal{V}}\right]_{T,n} 
   \nonumber \\
   &=& - \frac{1}{N^3} \sum_{\bf k} \frac{\partial E_{\bf k}}{\partial a^3}
         (2 n_{\bf k} - 1)
        + \frac{n}{2}\left( \frac{n}{2} - 1 \right)
        \left( \frac{\partial U}{\partial a^3}
        + 24 \frac{\partial V}{\partial a^3} \right) \;,
\label{P}
\end{eqnarray}
where
\begin{eqnarray}
  \frac{\partial E_{\bf k}}{\partial a^3}
  &=& \frac{\epsilon_{\bf k} - {\bar\mu}}{E_{\bf k}}
  \left[ \frac{\partial \epsilon_{\bf k}}{\partial a^3}
  + \frac{n}{2} \left( \frac{\partial U}{\partial a^3}
  + 24 \frac{\partial V}{\partial a^3} \right) \right] \;, \nonumber \\
  \frac{\partial \epsilon_{\bf k}}{\partial a^3}
  &=& \frac{2t}{3a^2}
  \sum_{i = x,y,z} k_i \cos\left(\frac{2\pi}{N}k_i\right) \;, \nonumber \\
  \frac{\partial U}{\partial a^3}
  &=& - \frac{1}{a^6} \left[ \left( V_c^{(0)} - 5\frac{V_c^{(2)}}{a^2} \right)
  - 3 \left( V_{\sigma}^{(0)} - 5 \frac{V_\sigma^{(2)}}{a^2} \right) \right]
  \;, \nonumber \\
  \frac{\partial V}{\partial a^3}
  &=& - \frac{5}{6a^8} V_c^{(2)} \; .
  \label{dvda3}
\end{eqnarray}

The double occupancy per site $D$ is expressed as
\begin{eqnarray}
  D &\equiv&  \frac{1}{N^3}\sum_i
  \langle {\hat c}_{i\uparrow}^\dagger{\hat c}_{i\uparrow}
  {\hat c}_{i\downarrow}^\dagger{\hat c}_{i\downarrow}\rangle
   = \frac{1}{N^3}\frac{\partial E}{\partial U} \nonumber \\
   &=& \frac{\Delta^2}{4U^2} + \frac{n^2}{4}
   - \frac{n}{N^3} \sum_{\bf k} \left( \epsilon_{\bf k}
  - {\bar \mu}\right) \frac{n_{\bf  k}^2}{T} e^{E_{\bf k}/T} \; .
  \label{d}
\end{eqnarray}

The kinetic energy per site $KE$ is written as
\begin{eqnarray}
  KE &\equiv& - \frac{t}{N^3}\sum_{\langle i, j \rangle, \sigma}
     \langle{\hat c}^\dagger_{i \sigma}{\hat c}_{j \sigma}\rangle
     + 6 \frac{t}{N^3} \sum_{i \sigma} \langle
     {\hat c}^\dagger_{i \sigma}{\hat c}_{i \sigma} \rangle
     =  \frac{t}{N^3} \frac{\partial E}{\partial t} \nonumber \\
  &=& \frac{1}{N^3} \sum_{\bf k} \frac{\left( \epsilon_{\bf k}
  - {\bar \mu}\right)
  \left( \epsilon_{\bf k} + 6t \right)}{E_{\bf k}}
  \left[ 2 n_{\bf k}\left( 1 - \frac{n_{\bf k}}{T}E_{\bf k}
  e^{E_{\bf k}/T} \right)  - 1 \right] + 6t \;.
  \label{kinetic}
\end{eqnarray}

\bigskip
\bigskip
\bigskip

\section{Density wave phase}
The Hamiltonian in the coordinate space is reduced to
\begin{eqnarray}
  {\hat H}
  &=& - t \sum_{\langle i,j \rangle \sigma}
           {\hat c}_{i \sigma}^\dagger {\hat c}_{j \sigma}
   + (6t+n{\bar U}/2) \sum_{i\sigma}
      {\hat c}_{i\sigma}^\dagger {\hat c}_{i\sigma} \nonumber \\
  & & + (U/2 + 4 \eta V) \delta\sum_{i\sigma}
         \cos(2\pi{\bf n}_i\cdot {\bf q}/N)
         {\hat c}_{i\sigma}^\dagger {\hat c}_{i\sigma}\nonumber\\
  & & - \left[ (U/4) (n^2 + \delta^2/2)
              + V (6 n^2 +\eta \delta^2)\right] N^3 \;,
  \label{freeCC}
\end{eqnarray}
where the parameter $\eta$ depends on the dimension of the density
wave $d$, $\eta = 3 - 2d$.  Here, we have used
\begin{eqnarray}
   \sum_{\langle i,j \rangle \sigma}(n_{j\uparrow}+n_{j\downarrow})
      {\hat c}_{i\sigma}^\dagger {\hat c}_{i\sigma}
 &\rightarrow&
    \sum_{\langle i,j \rangle \sigma}(n_{i\uparrow}+n_{i\downarrow})
                   {\hat c}_{i\sigma}^\dagger {\hat c}_{i\sigma}
   +\delta\sum_{i {\bf e} \sigma}
    \cos(2\pi({\bf n}_i+{\bf e}) \cdot {\bf q}/N)
           {\hat c}_{i \sigma}^\dagger {\hat c}_{i \sigma} \nonumber \\
 &\simeq& 6n\sum_{i\sigma}{\hat c}_{i\sigma}^\dagger {\hat c}_{i\sigma}
    +2\eta\delta\sum_{i \sigma}
    \cos(2\pi{\bf n}_i \cdot {\bf q}/N)
           {\hat c}_{i \sigma}^\dagger {\hat c}_{i \sigma}
\end{eqnarray}
and
\begin{eqnarray}
  \sum_{\langle i, j \rangle}(n_{i\uparrow}+n_{i\downarrow})
                             (n_{j\uparrow}+n_{j\downarrow})
  &\rightarrow&
  \sum_{\langle i, j \rangle}(n_{i\uparrow}+n_{i\downarrow})
                             (n_{j\uparrow}+n_{j\downarrow})
   +2\delta\sum_{\langle i j \rangle \sigma}
    \cos(2\pi{\bf n}_i \cdot {\bf q}/N)
                             (n_{j\uparrow}+n_{j\downarrow}) \nonumber \\
  & & +{\delta}^2\sum_{i {\bf e}}
         \cos({2\pi}{\bf n}_i\cdot {\bf q}/N)
         \cos({2\pi}({\bf n}_i+{\bf e})\cdot {\bf q}/N)
  \nonumber \\
  &\simeq&
    ( 6 n^2 +\eta {\delta}^2 ) N^3 \;,
\label{nnnn}
\end{eqnarray}
where the nucleon densities $n$'s are defined in the same way as
in Sec.~III. In the last step of Eq.~(\ref{nnnn}), we have used
the identities
\begin{equation}
  \begin{array}{l}
    \displaystyle{
      \sum_i \cos(2\pi{\bf n}_i \cdot {\bf q}/N)
      = 0 
    } \\
    \displaystyle{
      \sum_i \cos^2 (2\pi{\bf n}_i \cdot {\bf q}/N)
      = \frac{N^3}{2}
    } 
  \end{array}
\end{equation}
by applying the discretized orthonormality relation Eq.~
(\ref{deltafunc}).

Using
\begin{equation}
\sum_{i\sigma}\cos\left(\frac{2\pi}{N}{\bf n}_i \cdot {\bf q}\right) {\hat
c}_{i\sigma}^\dagger {\hat c}_{i\sigma} =\frac{1}{2}\sum_{{\bf
k}\sigma}\left(
                      {\hat c_{{\bf k}-{\bf q}/2 \; \sigma}}^\dagger
                       \hat c_{{\bf k}+{\bf q}/2 \; \sigma}
                     +{\hat c_{{\bf k}+{\bf q}/2 \; \sigma}}^\dagger
                       \hat c_{{\bf k}-{\bf q}/2 \; \sigma}
                      \right) \;,
\end{equation}
we obtain the momentum space representation
\begin{eqnarray}
  \hat{H} -\mu N^3 {\hat n}
  &=& \frac{1}{2}\sum_{{\bf k}\sigma}
       \left(\epsilon _{+} - {\bar \mu} \right)
       {\hat c}_{+ \sigma}^\dagger {\hat c}_{+ \sigma}
    + \frac{1}{2}\sum_{{\bf k}\sigma}
       \left(\epsilon _{-} - {\bar \mu} \right)
       {\hat c}_{- \sigma}^\dagger {\hat c}_{- \sigma} \nonumber \\
  & & + \frac{1}{8} {\bar \delta} \sum_{{\bf k} \sigma}
  \left( {\hat c}_{- \sigma}^\dagger {\hat c}_{+ \sigma}
      + {\hat c}_{+ \sigma}^\dagger {\hat c}_{- \sigma} \right)
  -\frac{1}{2}\delta {\bar \delta} N^3
      + E_0 N^3  \nonumber \\
  &=& \frac{1}{2}\sum_{{\bf k}\sigma}
  \left( {\hat c}_{+ \sigma}^\dagger
  \;\;\; {\hat c}_{- \sigma}^\dagger \right)
  \left(
  \begin{array}{cc}
  \epsilon_{+}-{\bar \mu} &  \frac{1}{4}{\bar \delta} \\
  \frac{1}{4}{\bar \delta}  &  \epsilon_{-}-{\bar \mu}
  \end{array}
  \right)
  \left(
  \begin{array}{c}
  {\hat c}_{+ \sigma} \\
  {\hat c}_{- \sigma}
  \end{array}
  \right)  \nonumber \\
  & & -\frac{1}{16}\delta {\bar \delta} N^3
      + E_0 N^3  \;,
  \label{DWgscfH}
\end{eqnarray}
where the subscripts $+$ and $-$ denote ${\bf k}+{\bf q}/2$ and
${\bf k}-{\bf q}/2$, respectively; that is, ${\hat c}_{\pm \sigma}
= {\hat c}_{{\bf k}\pm{\bf q}/2 \; \sigma}$. $\epsilon_{\pm}$ is
defined as
\begin{displaymath}
  \epsilon_{\pm} = - t \sum_{\bf e}
  \exp \left( i 2\pi
  ({\bf k}\pm{\bf q}/2) \cdot {\bf e} /N\right)
  = -2t \sum_{j=x,y,z} \cos \left(2\pi(k_j \pm q_j/2)/N\right)
  \; .
\end{displaymath}
We have also defined
\begin{equation}
 {\bar \delta} = \delta \left( 2U +16\eta V \right) \;.
\end{equation}

${\hat H} -\mu N^3 {\hat n}$ is diagonalized through the
transformation
\begin{equation}
  \begin{array}{l}
  \displaystyle{
    {\hat \alpha}_{{\bf k}+}
    = u_{{\bf k}} {\hat c}_{+ \sigma}
    + v_{{\bf k}} {\hat c}_{- \sigma} \;, 
  } \\
  \displaystyle{
    {\hat \alpha}_{{\bf k}-}
    = u_{{\bf k }} {\hat c}_{- \sigma}
    - v_{{\bf k}} {\hat c}_{+ \sigma} \;, 
  }
  \end{array}
\end{equation}
where $u_{{\bf k}}$ and $v_{{\bf k}}$ are taken to be real and are
given by
\begin{equation}
  u_{\bf k}^2 = \frac{1}{2} \left( 1 +
  \frac{\epsilon_+ - \epsilon_-}{\xi_{\bf k}} \right)
  \;\;\;{\rm and} \;\;\;
  v_{\bf k}^2 = \frac{1}{2} \left( 1 -
  \frac{\epsilon_+ - \epsilon_-}{\xi_{\bf k}} \right)
    \; ,
  \nonumber
\end{equation}
satisfying $u_{\bf k}^2 + v_{\bf k}^2 = 1$. Here,
\begin{equation}
\xi_{\bf k} = \sqrt{(\epsilon_+-\epsilon_-)^2
           +{\bar \delta}^2/4 }  \;.
\end{equation}

${\hat H} -\mu N^3 {\hat n}$ is now expressed as that of a system
of free quasi-particles:
\begin{equation}
{\hat H} -\mu N^3 {\hat n} = \frac{1}{2}\sum_{{\bf k}\lambda}
  E_{{\bf k}\lambda}
  {\hat \alpha}_{{\bf k}\lambda}^\dagger {\hat \alpha}_{{\bf k}\lambda}
  + N^3 (E_{\rm GS} - \mu n) \; ,
  \label{Hcannew}
\end{equation}
where $\lambda =\pm$ and
\begin{equation}
  E_{{\bf k}\pm} = \frac{1}{2}\left[
     (\epsilon_++\epsilon_--2{\bar \mu}) \pm \xi_{\bf k}\right]
\end{equation}
is the energy of the quasi-particles, and
\begin{equation}
  E_{\rm GS} = - \delta {\bar \delta}/16
               - n^2 {\bar U}/4
               + \mu n
  \label{Egsnew}
\end{equation}
is the ground-state energy of the system.

\vfill\eject


\vfill\eject


\begin{thebibliography}{99}
\bibitem{dean}D. J. Dean and M. Hjorth-Jensen, Rev. Mod. Phys.
{\bf 75}, 607 (2003), and references quoted therein.
\bibitem{arpns} H. Heiselberg and V. R. Pandharipande, Annu. Rev. Nucl.
Part. Sci. {\bf 50}, 481 (2000), and references quoted therein.
\bibitem{stat} For example, A. Bohr and B. R. Mottelson, {\it Nuclear
Structure} (World Scientific, Singapore, 1998), Vol. I, pp. 179 --
188.
\bibitem{latgas} C. B. Das, S. Das Gupta, and S. K. Samaddar, Phys. Rev.
C {\bf 63}, 011602 (2001) and references therein; J. Borg, I. N.
Mishustin, and J. P. Bondorf, Phys. Lett. B {\bf 470}, 13 (1999); X.
Campi and H. Krivine, Nucl. Phys. {\bf A620}, 46 (1997); and T. T.
S. Kuo, S. Ray, J. Shamanna, and R. K. Su, Int. J. Mod. Phys. E
{\bf 5}, 303 (1996).
\bibitem{TTT} J.-R. Buchler and S. A. Coon, Astrophys. J. {\bf 212},
807 (1977).
\bibitem{panda-t} B. Friedman,
V. R. Pandharipande, and Q. N. Usmani, Nucl. Phys. {\bf A372}, 483
(1981); B. Friedman and V. R. Pandharipande, {\it ibid}. {\bf 
A361}, 502 (1981).
\bibitem{kyoto-t}
T. Takatsuka, Prog. Theor. Phys. {\bf 80}, 475 (1989); T.
Takatsuka and J. Hiura, {\it ibid}. {\bf 79}, 268 (1988).
\bibitem{other-t}
H. M\"{u}ther and A. Sedrakian, Phys. Rev. C {\bf 67}, 015802
(2003); A. Schnell, G. R\"{o}pke, and P. Schuck, Phys. Rev. Lett.
{\bf 83}, 1926 (1999); M. Baldo, J. Cugnon, A. Lejeune, and U.
Lombardo, Nucl. Phys. {\bf A515}, 409 (1990); R.-K. Su and F.-M.
Lin, Phys. Rev. C {\bf 39}, 2438 (1989); R. K. Su, S. D. Yang, and
T. T. S. Kuo, {\it ibid}. {\bf 35}, 1539 (1987); H. R.
Jaqaman, A. Z. Mekjian, and L. Zamick, {\it ibid}. {\bf 29}, 2067
(1984).
\bibitem{jkuo}M. F. Jiang and T. T. S. Kuo, Nucl. Phys. {\bf A481}, 294
(1988).
\bibitem{gapref} {\O}. Elgar{\o}y, L. Engvik, M. Hjorth-Jensen,
and E. Osnes, Nucl. Phys. {\bf A604}, 466 (1996).
\bibitem{panda}
J. Morales, Jr., V. R. Pandharipande, and D. G. Ravenhall, Phys.
Rev. C {\bf 66}, 054308 (2002); A. Akmal, V. R. Pandharipande, and
D. G. Ravenhall, {\it ibid}. {\bf 58}, 1804 (1998); A. Akmal and
V. R. Pandharipande, {\it ibid}. {\bf 56}, 2261 (1997); R. B.
Wiringa, V. Fiks, and A. Fabrocini, {\it ibid}. {\bf 38}, 1010
(1988).
\bibitem{kyoto}
T. Muto, T. Takatsuka, R. Tamagaki, and T. Tatsumi, Prog. Theor.
Phys. Suppl. {\bf 112}, 221 (1993); T. Kunihiro, T. Muto, T.
Takatsuka, R. Tamagaki, and T. Tatsumi, {\it ibid}. {\bf 112}, 307
(1993), and references therein.
\bibitem{fantoni} K. E. Schmidt and S. Fantoni, Phys. Lett. B {\bf
446}, 99 (1999); A. Sarsa, S. Fantoni, K. E. Schmidt, and F.
Pederiva, Phys. Rev. C {\bf 68}, 024308 (2003), and references
therein.
\bibitem{MKSV} H.-M. M\"{u}ller, S. E. Koonin, R. Seki, and U. van Kolck, Phys.
Rev. C {\bf 61}, 044320 (2000).
\bibitem{RMP:MRR} R. Micnas, J. Ranninger, and S. Robaszkiewicz, Rev. Mod.
Phys. {\bf 62}, 113 (1990);  J. E. Hirsch and D. J. Scalapino, Phys. Rev.
B {\bf 27}, 7169 (1983); 
N. Andrenacci, A. Perali, P. Pieri, and G. C. Strinati, {\it ibid}.
{\bf 60}, 12410 (1999), and references therein.
\bibitem{Fradkin} E. Fradkin, {\it Field Theories of Condensed Matter
Systems} (Addison-Wesley, Redwood City, 1991), Chap. 5.
\bibitem{tara} For example, A. Taraphder, H. R. Krishnamurthy, R.
Pandit, and T. V. Ramakrishnan, Phys. Rev. B {\bf 52}, 1368
(1995).
\bibitem{HFB} P. Ring and P. Schuck, {\it The Nuclear Many-Body Problem}
(Springer-Verlag, Berlin, 1980) pp. 244--258; J.-P. Blaizot
and G. Ripka, {\it Quantum Theory of Finite Systems} (MIT Press,
Cambridge, Mass., 1986) pp. 193--204; J. G. Valatin, Phys. Rev.
{\bf 122}, 1012 (1961).
\bibitem{Leg} A. J. Leggett, J. Phys. (Paris) Colloq. {\bf 41}, C7--19
(1980).
\bibitem{Noz} P. Nozieres and S. Schmitt-Rink, J. Low Temp. Phys. {\bf
59}, 195 (1985).
\bibitem{schrieffer} For example, J. R. Schrieffer, {\it Theory of
Superconductivity} (Benjamin, New York 1964), p. 55.
\bibitem{renorm} N. Goldenfeld, {\it Lectures on Phase Transitions
and the Renormalization Group} (Perseus Books, Reading, Mass.,
1992), p. 153, and references quoted therein.
\bibitem{armen} A. N. Kocharian, C. Yang, Y. L. Chiang, Phys. Rev.
B {\bf 59}, 7458 (1999).
\bibitem{MS} H.-M. M\"{u}ller and R. Seki, in {\it Nuclear Physics with
Effective Field Theory}, edited by R. Seki, U. van Kolck, and M.
J. Savage (World Scientific, Singapore, 1998), pp. 191--197.
\bibitem{lstv} D. Lee, R. Seki, R. Timmerman, and U. van Kolck, (unpublished).
\bibitem{pb} T. Papenbrock and G. F. Bertsch, Phys. Rev. C {\bf 59},
2052 (1999).
\bibitem{NN}  For example, A. Bohr and B. R. Mottelson, {\it Nuclear
Structure} (World Scientific, Singapore, 1998), Vol. I, pp. 263 --
268.
\bibitem{TT} T. Takatsuka and R. Tamagaki, Prog. Theor. Phys. Suppl.,
No. 112, 27 (1993).
\bibitem{superc}
R. Meservey and B. B. Schwartz, in {\it Superconductivity}, edited by R.
D. Parks (Marcel Dekker, N.Y., 1969) pp. 117--191.
\bibitem{schwenk} A. Schwenk, B. Friman, and G. E. Brown, Nucl.
Phys. {\bf A713}, 191 (2003).
\bibitem{skyrme} T. H. R. Skyrme, Philos. Mag. {\bf 1}, 1043 (1956);
Nucl. Phys. {\bf 9}, 615 (1959).
\bibitem{luescher} M. L{\"u}scher, Commun. Math. Phys. {\bf 105},
153 (1986); Nucl. Phys. B {\bf 354}, 531 (1991).
\bibitem{wash} S. R. Beane, P. F. Bedaque, A. Parreno, and M. J.
Savage, Phys. Lett. B {\bf 585}, 106 (2004).


\end{thebibliography}
\end{document}